\documentclass[aps,prl,amsmath,amssymb,twocolumn, nofootinbib]{revtex4}
\usepackage{color}

\usepackage{graphicx}
\usepackage{bm}
\usepackage{lipsum}
\usepackage{epsfig}
\usepackage{epstopdf}
\usepackage{colordvi}
\usepackage{float}
\usepackage{accents}

\begin{document}

\title{Resonance in modulation instability from non-instantaneous nonlinearities}
\author{Ray-Ching Hong,$^1$ Chun-Yan Lin,$^1$ You-Lin Chuang,$^{2}$ Chien-Ming Wu,$^{1}$ Yonan Su,$^{1}$\\ Jeng Yi Lee,$^{1}$ Chien-Chung Jeng,$^{3}$ Ming-Feng Shih,$^{4}$ and  Ray-Kuang Lee$^{1,2,5}$ }

\affiliation{
$^1$Institute of Photonics Technologies, National Tsing Hua University, Hsinchu 30013, Taiwan\\
$^2$Physics Division, National Center for Theoretical Sciences, Hsinchu 30013, Taiwan\\
$^3$Department of Physics, National Chung-Hsing University, Taichung 402, Taiwan\\
$^4$Department of Physics, National Taiwan University, Taipei 106, Taiwan\\
$^5$Department of Physics, National Tsing Hua University, Hsinchu 30013, Taiwan}

\begin{abstract}
To explore resonance phenomena in the nonlinear region, we show by experimental measurements and theoretical analyses that resonance happens in modulation instability (MI) from non-instantaneous nonlinearities  in photorefractive crystals. 
With a temporally periodic modulation in the external bias voltage, corresponding to a modulation in the nonlinear strength,  an enhancement in the visibility of MI at resonant frequency  is reported through spontaneous optical pattern formations.
Modeled by such temporally periodic nonlinear driving force to the system, theoretical curves obtained from a nonlinear non-instantaneous Schr\"{o}dinger equation give good agreement to experimental data.
As MI is a universal signature of symmetry-breaking phenomena, our observation on the resonance in MI may provide a control on chaotic, solitary, and turbulence waves.
\end{abstract}

\maketitle

As a simple means to observe the manifestation of strongly nonlinear effects in nature, modulation instability (MI) has played an important role in a variety of nonlinear systems, from sand ripples, cloud formations, water waves, to animal pigmentation~\cite{Bespalov, Karpman, fluid}. 
Such a symmetry-breaking phenomena driven through the stochastic fluctuations is  closely related to the pattern formations in optics, plasma physics, and hydrodynamics~\cite{noise-pattern}. 
In optics, MI causes  chaotic, solitary, or turbulence waves, whereby  a small perturbation in the amplitude or phase in the input wave grows exponentially~\cite{MI3, Agrawal}. 
Even though MI had been reported with biased photorefractive crystals two decades ago~\cite{Carvalho}, until recently the existence of optical pattern transitions, as well as their phase boundaries, in spontaneous optical pattern formations are observed~\cite{PRL09, boundary}. 

As the input noises being amplified and optimized, MI  also accounts for stochastic resonance~\cite{SR, noisyimage, SR-MI}, competition and correlation in confined patterns ~\cite{nphoton},  degradation of beam quality in high power laser systems~\cite{fiberlaser, D3, twin-core}, and emergence of giant rogue wave/super-continuum generation~\cite{rogue1, rogue2, SC}. 
In addition to being  an intrinsic property in nonlinear systems, significant interest grows in the suppression or modification of MI, such as by means of partial coherent light~\cite{Kip, Soljacic, Dylov}, nonlocal nonlinear media~\cite{nonlocal},  periodically tapered photonic crystal fibers~\cite{tunable}, or the intensity ratio between background to signal fields~\cite{ratio}.

\begin{figure}[b]
\centering
\includegraphics[width=8.4cm, height=4.5cm]{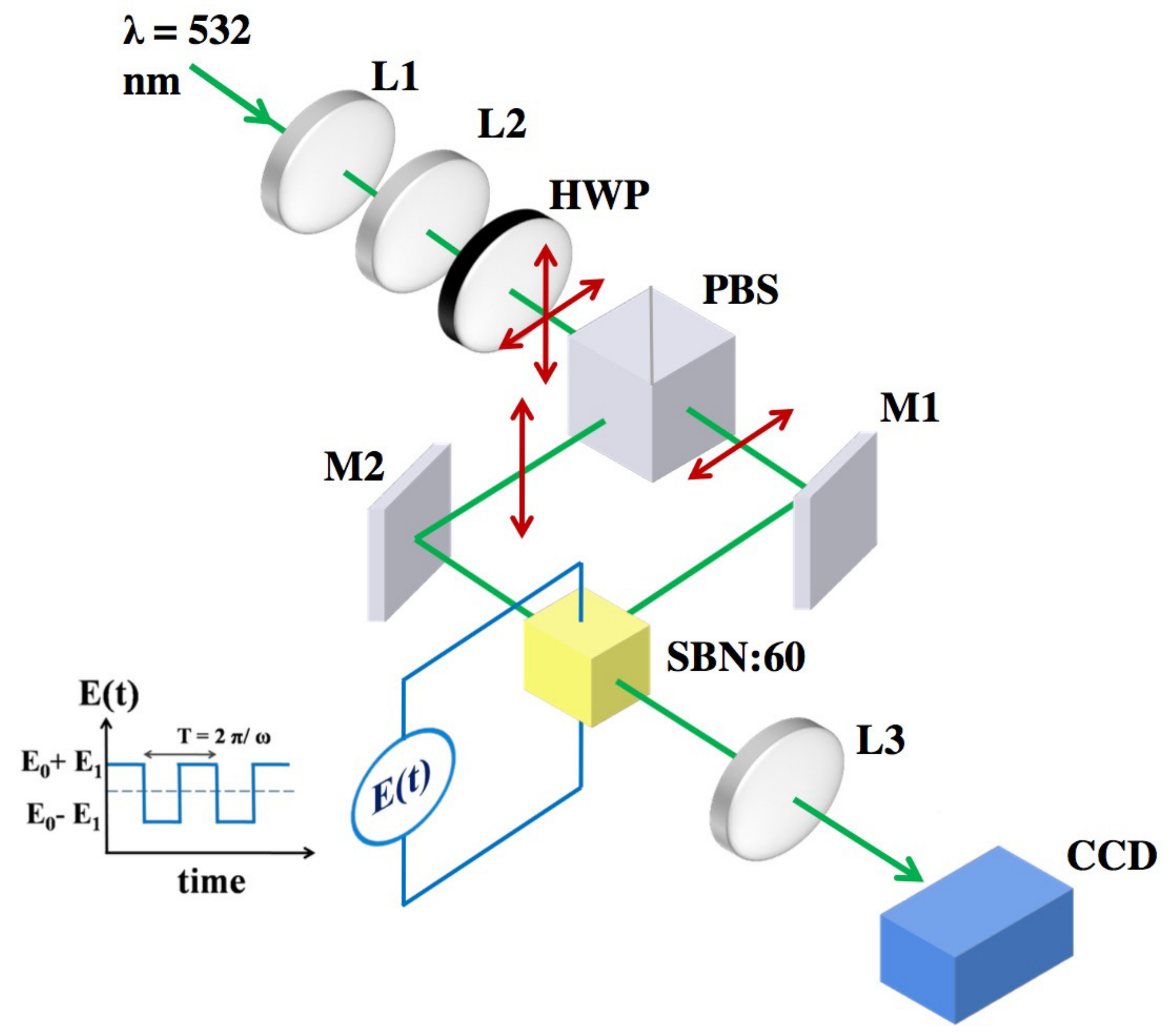}
\caption{Schematic diagram for our experimental setup, where a laser light source at the wavelength $\lambda = 532$ nm is incident onto a photorefractive crystal, SBN:60. Here, L1 and L2 are two planoconvex lenses;  HWP denotes a half-wave plate; PBS is the polarization beam splitter; M1 and M2 are two reflecting mirrors; and  L3 is an imaging lens to collect optical patterns into a CCD camera. An external bias voltage E(t) is applied to the crystal, with a periodically temporal modulation in a square wave function (biased at E$_0$, with the modulation depth E$_1$,  and period T=$2\pi/\omega$).}
\end{figure}

In this Letter,  by experimental measurements and theoretical analyses we demonstrate a directly temporal modulation in nonlinearities through a periodic change in the external bias voltage.
When driven by an external force or by varying some parameters of the system, a universal phenomenon, resonance, may happen with greater amplitude in the output  at some frequency.
Unlike previous work only introducing linear periodic driving forces, we report nonlinear resonance by predicting and observing MI with a greater visibility at specific frequencies.
With spontaneous optical pattern formations of MI in photorefractive crystals,  enhancement in the visibility is achieved due to the spatialtemporal response in non-instantaneous nonlinearities~\cite{MI88,vortex}.
The results demonstrated here extend our understandings on resonant phenomena, with applications to control chaotic, solitary, and turbulence waves.

\begin{figure*}[ht]
\includegraphics[width=17.0cm]{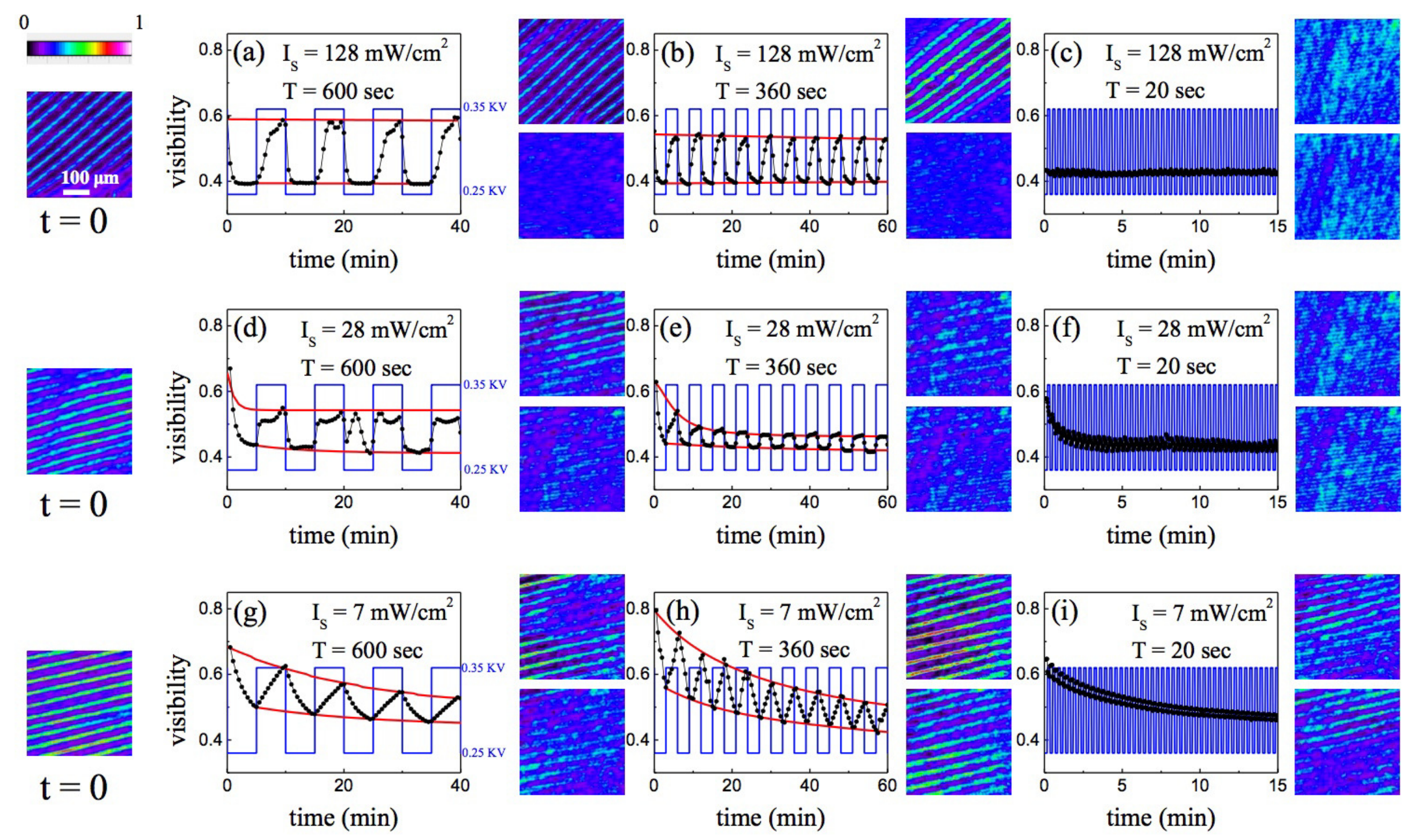}
\caption{Visibility of MI and the corresponding optical intensity patterns collected at the output plane through a non-instantaneous photorefractive crystal.
Here, we apply a bias voltage modulated in a square wave function (E$_0 = 0.3$kV and E$_1=0.05$kV) but with different periods, i.e., T$=600$, $360$, and $20$ seconds for the left, middle, and right columns.
Three different signal intensities (with the initial MI patterns at $t = 0$ shown in the leftest panel accordingly) are used, i.e., I$_s = 128$, $28$, and $7$ mW/cm$^2$ from top to bottom rows, which give instantaneous, intermediate, and non-instantaneous nonlinear responses, respectively.
In each panel, the visibility of MI pattern is recorded as a function of time; while two images shown in false colors are collected at the maximum (upper) and minimum (lower) biased voltages E$(t) =$ E$_0\pm$ E$_1$. }
\end{figure*}

The schematic digram for our experiment setup is shown in Fig. 1, where a Nd:YVO$_4$ diode-pumped, continuous wave, solid state laser at the center wavelength  $\lambda=532$ nm is used as the input light.
Then, we use two planoconvex lenses L1 and L2 for beam collimation.
The signal beam ($o$-wave) and background beam (e-wave), maintaining at a constant ratio $4:1$, are controlled by half-wave plate (HWP) and polarization beam splitter (PBS).
The nonlinearity in our system is provided by a photorefractive crystal, i.e., a strontium-barium niobate (SBN:60) crystal, which has $5\times 5 \times 5$ mm$^{3}$ in size, along with an effective electro-optical coefficient $r_{33} = 350$ pm/V.
In the output plane, the image of signal beam is collected by a charge-coupled device (CCD) camera. 
Contrast to our previous wok~\cite{MI88, PRL09, boundary, ratio, vortex}, instead of using a constant bias voltage to fix the nonlinear strength, here, we  apply  a square wave function on the voltage, in order to  give a periodic change in the nonlinear strength.

For fields propagating in a SBN crystal~\cite{band-transport, band-transport2, t-pattern}, we can apply the nonlinear Schr\"{o}dinger equation  with a non-instantaneous nonlinear response characterized by a time constant $\tau$ in the photorefractive system, i.e.,
\begin{equation}
 i\frac{\partial A}{\partial z}+\frac{1}{2}\frac{\partial^2 A}{\partial x^2}-\int_{-\infty}^{t}\texttt{d}t_1 [ \frac{\gamma(t_1)}{\tau}F(|A(t_1)|^2)e^{-(t-t_1)/\tau}] A = 0.
\label{eq:eqSE}
\end{equation}
Here, the spatial coordinates are normalized to $1/k_0$ with $k_0 = 2\pi/\lambda$ being the wavenumber of incident field, and $A$ is normalized to $\sqrt{I_s}$ given that $I_s$ being the saturation intensity (as the  background beam).
The nonlinear response function is given as $F(|A|^2)=1/(1+|A|^2)$; while the nonlinear strength
$\gamma (t) \propto r_{33}\, \text{E}(t) \equiv \gamma_0 \, \text{E}(t)$ is assumed linearly proportional to the bias voltage E$(t)$, but with a periodic change in time.

In Fig. 2, we demonstrate series of optical patterns recorded by the CCD camera at the output plane when the bias voltage is modulated by a square wave function with DC term E$_0$,  modulation depth E$_1$,  modulation frequency $\omega = 2\pi f$, and period T=$2\pi/\omega$ .
When biased at E$_0 = 0.3$ kV, our photorefractive crystal is operated above the threshold voltage for MI pattern to emerge~\cite{ratio}.
After reaching a steady output in the spontaneous pattern formation, as shown in the leftest panel of each row in Fig. 2, we set the time as $t = 0$ and start to modulate the external bias voltage with a square wave function.
Here, we fix the modulation depth E$_1 = 0.05$ kV, but vary the time period in the applied square wave function, i.e., T$=600$, $360$, and $20$ seconds, as shown in the left, middle, and right columns of Fig. 2.
Moreover, as the nonlinearity  in photorefractive crystal is relaxed with a time constant of non-instantaneous response~\cite{timeconstant}, we also apply three different signal intensities in the inputs, i.e., I$_s = 128$, $28$, and $7$ mW/cm$^2$ from top to bottom rows, in oder to illustrate nearly instantaneous, intermediate, and non-instantaneous nonlinear responses, respectively.

For quantitatively analyses, here, each MI pattern is characterized by its visibility in the optical image, defined as V$\equiv (I_{\text{max}}-I_{\text{min}})/(I_{\text{max}}+I_{\text{min}})$ with $I_{\text{max}}$ and $I_{\text{min}}$ referring to the maximum and minimum values obtained from the optical intensity patterns in CCD.
With a long period in the modulation, such as T$=600$ seconds shown in Fig. 2(a), one can see that the visibility in the corresponding MI patterns vary in a periodic way  with the same oscillation frequency as that in the applied biased voltage.
Two typical images collected at the maximum  and minimum  biased voltages E$ = 0.3 \pm 0.05$ are revealed accordingly.
Experimentally, such a  periodic oscillation in the MI pattern can be observed even  for a very long time (more than one hour).

Then, we increase the modulation frequency $\omega$ in the applied bias voltage, or equivalently reduce the period T.
When the period is reduced to T$= 360$ seconds, as shown in Fig. 2(b), one can also see a periodic change in the visibility of MI patterns, which sustains up to $60$ minutes, too.
Two curves in Red color describing the envelop in the recording curve of visibility are depicted to indicate the supported range in visibility.
We refer the difference in the visibilities, or equivalently the distance between these two Red curves as the {\it band for visibility}, denoted as $\Delta$ in the following.
Compared to the scenario with a longer period shown in Fig. 2(a), the supported band for visibility also decreases.
Moreover, with a very short period in the change of biased voltage,  it is difficult to see the change in the resulting visibility of MI, as shown in Fig. 2(c) with T$= 20$ seconds. 
With the images shown in Figs. 2(a-c), naively, we can say that the visibility of MI patterns follows the change in the applied biased voltage when the modulation frequency is small.
However, when the speed of change in the external modulation voltages is too high,  the nonlinear response in our photorefractive crystal can not follow, resulting in the output visibility maintaining at a constant value.

When the input signal intensity is reduced to I$_s = 28$ mW/cm$^2$, similar scenarios can be seen in the middle row of Fig. 2.
However, as one can see in Figs. 2(d-f), now we have a longer relaxation time due to a weaker input signal, which causes a delayed response in the visibility of MI pattern, resulting in a {\it damped oscillation}.
Such a damped oscillation may come from the relaxation of quasi-static electric fields in photorefractive crystals~\cite{dynamics, quasi-static}.
Moreover, the supported band for visibility in MI patterns $\Delta$ shrinks when the  modulation period is reduced from T$= 600$, $360$, to $20$ seconds.

\begin{figure}[b]
\centering
\includegraphics[width=4.2cm]{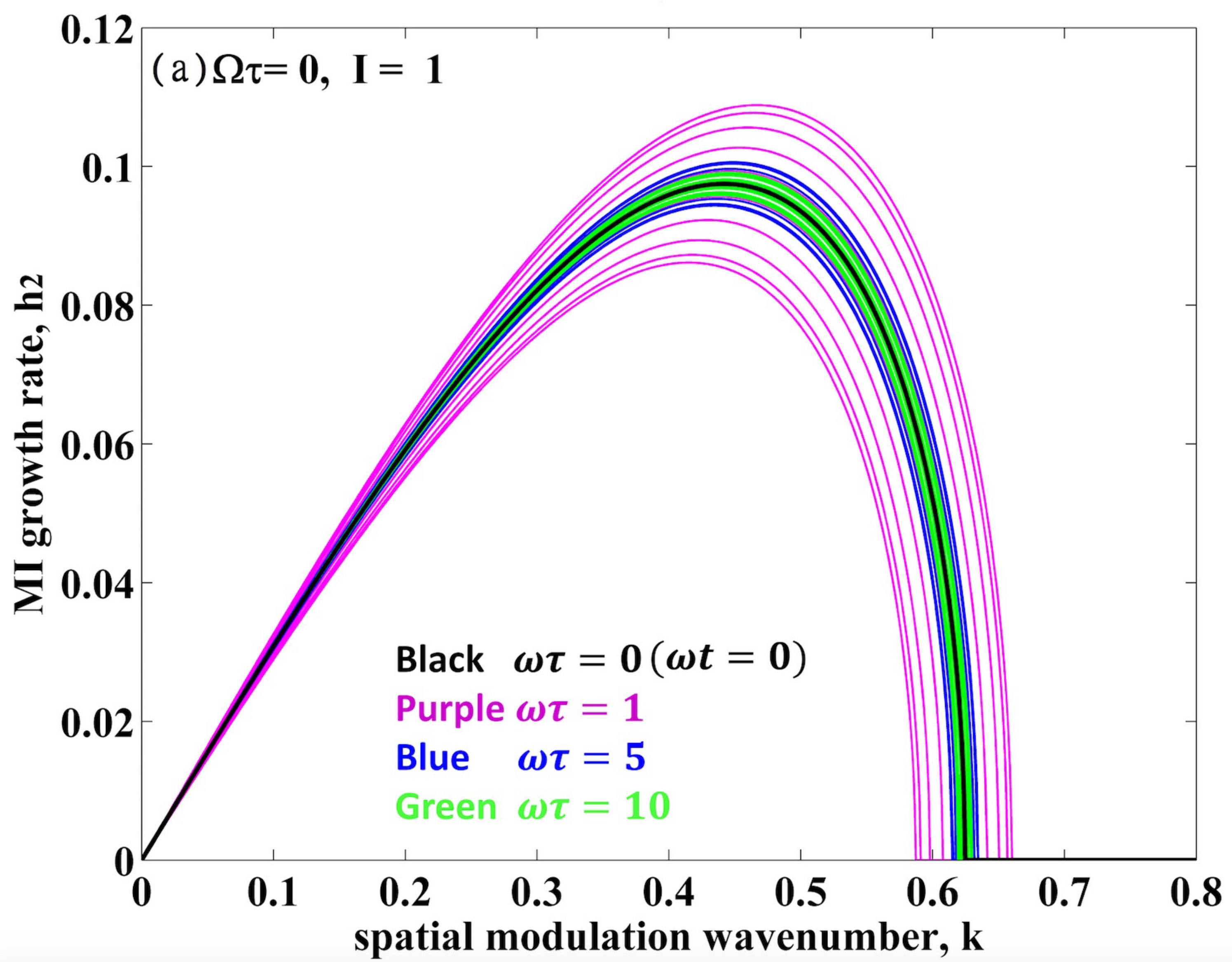}
\includegraphics[width=4.2cm]{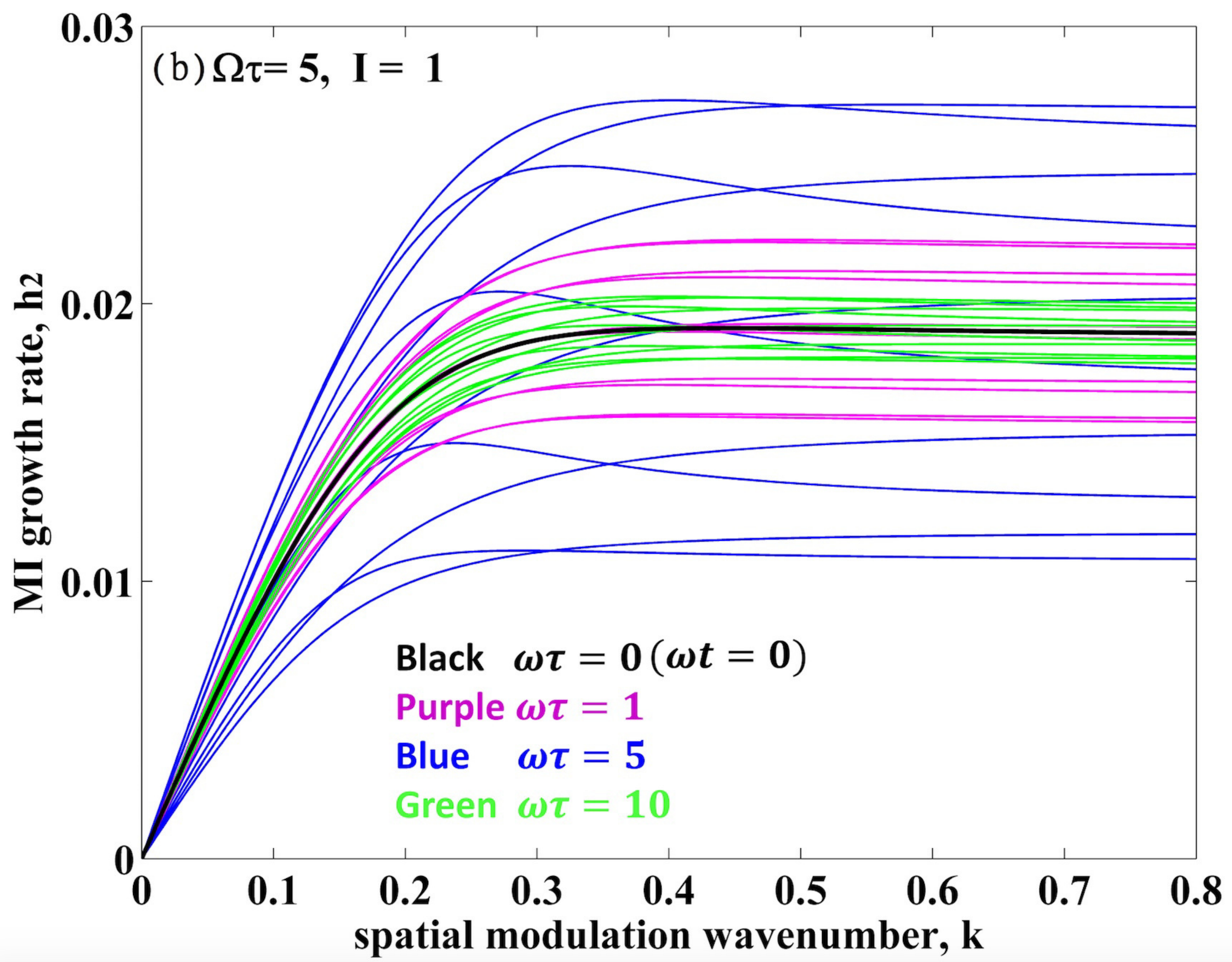}
\caption{MI spectrum for  nonlinear systems with (a) instantaneous ($\Omega \tau = 0$) and (b) non-instantaneous ($\Omega \tau = 5$) responses. Different external modulation frequencies are depicted for $\omega \tau = 0$, $1$, $5$, and $10$ in Black, Purple, Blue, and Green colors, respectively; along with (in the same colors) $\omega t = n\times 2\pi/10$ ($n = 0, 1, \cdots 9$).
Other parameters used are the normalized input field intensity I $\equiv |A|^2 = 1$, E$_0 = 0.3$, E$_1 = 0.05$, and nonlinear factor $\gamma_0 = 1.3$, respectively.}
\end{figure}

When one further decreases the signal intensity to $7$ mW/cm$^2$, as shown in the third row of Fig. 2, the relaxation time constant in the nonlinear response becomes  significantly   long enough.
Now,  our photorefractive crystal gives a non-instantaneous nonlinear response.
Again, a damped oscillation can be observed in the visibility of MI pattern, see Figs. 2(g-i).
When the modulation period decreases, on the contrary to the reduction in the supported band in visibility, a great enlargement in MI visibility can be clearly seen when we move the modulation period from T$= 600$ to $360$ seconds as shown in Figs. 2(g) and (h).
Then, when the modulation period goes smaller,  the supported band in visibility becomes a narrow one.

To explain these  experimental images of spontaneous optical pattern formations, we illustrate the underly picture by finding the corresponding resonant frequencies when one modulates the nonlinear strength periodically.
Without loss of generality, we assume that a sinusoidal wave function is applied to the  bias voltage E$(t)$ given in Eq. (1), i.e.,  $\text{E}(t)= \text{E}_0 + \text{E}_1 \,\text{Sin} (\omega t)$, with the DC term E$_0$,  AC amplitude E$_1$,  and modulation frequency $\omega = 2\pi/T$ in the external periodic function.

The corresponding MI spectrum for the unstable perturbed field on top of a plane wave solution can be found by applying $A=[A_{0}+a(x,z,t)]\exp[-i\beta z]$ into Eq. (1), with an assigned propagation constant $\beta$. 
By casting the perturbed field in the form of $a = [a_0\exp(i \Omega t + i k x - i h z) + c.c.]$, with the temporal modulation frequency $\Omega$ and spatial wavenumber $k$, respectively, one can determine an unstable perturbed field as its perturbed propagation constant is no longer a real number, i.e., $h = h_{1}+i\, h_{2}$.
The growth rate of this unstable perturbed field is referred as $h_2$, which has the form
\begin{widetext}
\begin{eqnarray}
h_2(k, I,\Omega\tau,\omega\tau, t)=
\pm \text{Re}\left[\sqrt{\frac{-k^4}{4}+
\frac{\gamma_0\, k^2\, |A|^2}{(1+|A|^2)^2}\left\{ \frac{\text{E}_0}{1+i\, \Omega \tau} + \frac{\text{E}_1}{2\, i}\left[\frac{\text{exp}[i\omega t]}{1+i (\Omega+\omega) \tau}-\frac{\text{exp}[-i\omega t]}{1+i (\Omega-\omega) \tau}\right]\right\}
}\,\right].
\end{eqnarray}
\end{widetext}
When the modulation frequency $\omega = 0$, the growth rate shown in Eq. (2) is reduced to the spatiotemporal MI of coherent light in non-instantaneous nonlinear media~\cite{MI88}, whereby the corresponding growth rate spectrum becomes a flat one at​ ​higher frequencies​.​

However, with a time-dependent modulation in the nonlinear strength, $\omega \neq 0$, the corresponding growth rate varies accordingly with the same frequency.
In Fig. 3, based on Eq. (2), we reveal the MI spectrum, in terms of the MI growth rate $h_2$ as a function of the spatial  wavenumber $k$ both for instantaneous and non-instantaneous nonlinearities, but with an external modulation in the nonlinear strength.
Let us consider the case with instantaneous nonlinearity first, i.e., $\Omega \tau \ll 1$.
As depicted in Fig. 3(a) for $\Omega \tau = 0$, instead of a fixed value (see the curve in Black color for $\omega \tau = 0$), there exists a band for the  MI growth rate $h_2$ to vary when the nonlinear strength is modulated at $\omega$.
Such a time-dependent growth rate can vary within a wider band region as the modulation frequency $\omega$ is small (but non-zero), which indicates that a change in the nonlinearity provided by photorefractive crystals just follows the change in bias voltage.
Nevertheless, when the modulation frequency increases, the size of supported band for the growth rate shrinks, as shown in Fig. 3(a) for $\omega \tau = 1$, $5$, and $10$ in Purple, Blue and Green colors, respectively, similar to the observation in the experiment shown in the first row of Fig. 2.
As the nonlinear response in photorefractive crystals comes from Pockel  effect through the transport of carrier donors, when $\omega \gg \Omega$, a fast periodic modulation in the applied voltage can only give an average value of the nonlinear strength, i.e., 
$h_2(\Omega\tau=0,\omega\tau\rightarrow \infty, t) \approx \pm \text{Re}\left[\sqrt{\frac{-k^4}{4}+
\frac{\gamma_0\, k^2\, |A|^2}{(1+|A|^2)^2}\text{E}_0} \right]$.
It means that we have a single value in the MI growth rate supported by nothing but the DC term ($\omega = 0$) when the modulation frequency is high enough.

However, the scenario is totally different for a non-instantaneous nonlinear system, i.e., $\Omega \tau \neq 0$.
As shown in Fig. 3(b), even though there exist several maximum values to support MI (experimentally only the lower value survives due to the spatial diffraction), the corresponding MI profile changes with a non-zero external modulation frequency $\omega\tau \neq 0$.
Nevertheless, as one can see from Eq. (2), resonance can happen when the external modulation frequency approaches the temporal modulation frequency in MI, i.e.,  $\omega \rightarrow \Omega$.
In particular, one can see clearly that the corresponding gain profile changes within a very large range, as shown in Blue color,  when the external modulation frequency is  the same as that in the non-instantaneous nonlinear response, i.e., $\omega = \Omega$.

\begin{figure}[t]
\centering
\includegraphics[width=8.4cm, height=6.0cm]{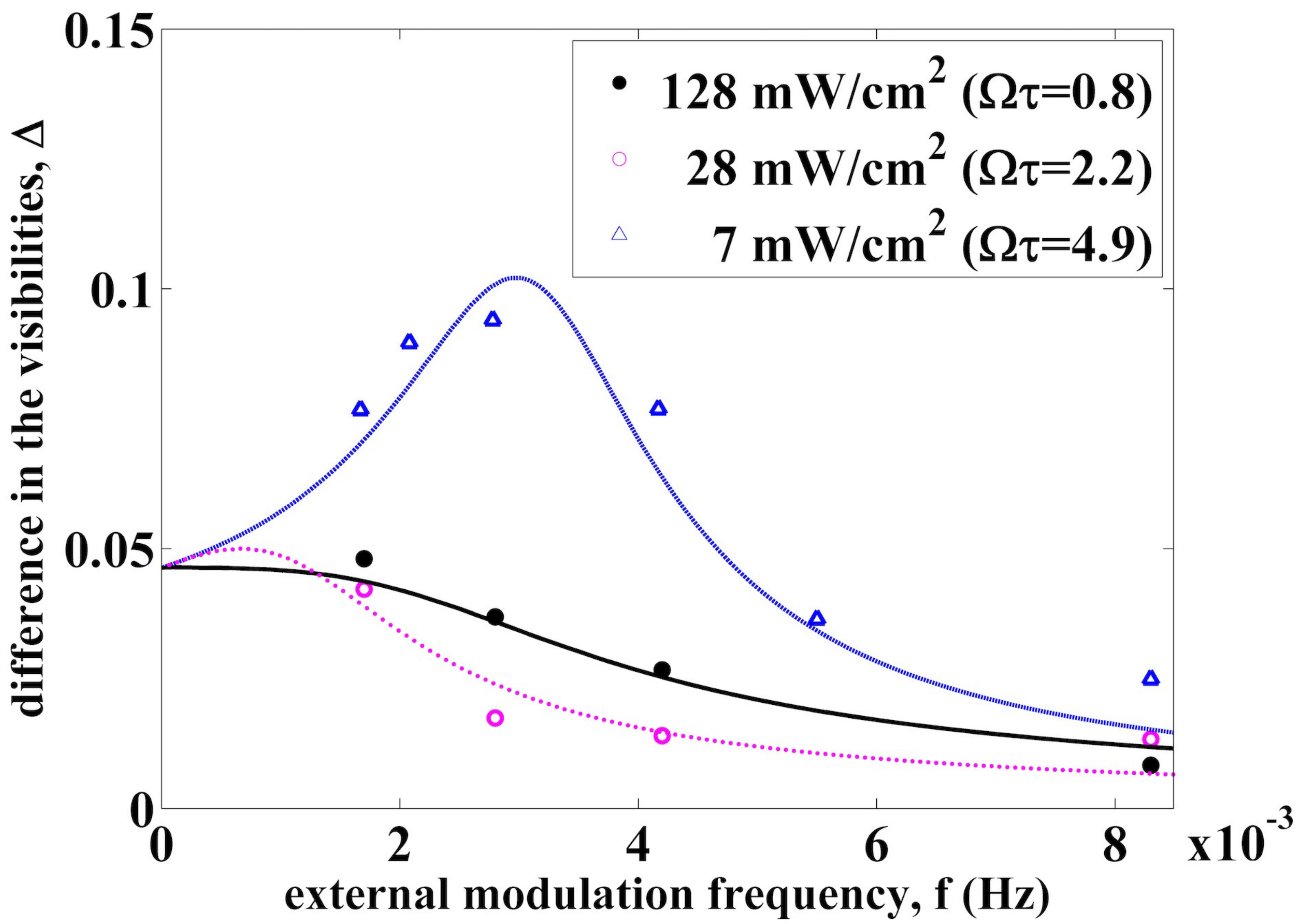}
\caption{Spectrum for MI visibility, defined through the  difference in the visibilities $\Delta$ as a function of external modulation frequency $f$ (Hz). Here, marked points are experimental data; while theoretical curves  are obtained from Eq. (2).
Three different relaxation time constants, i.e., $\Omega \tau = 0.8$, $2.2$, $4.9$  are fitted for instantaneous, intermediate, and non-instantaneous nonlinear responses, respectively.
Note that all the values at $f = 0$ are normalized to the same value.}
\end{figure}

With the theoretical growth rate given in Eq. (2),  we plot the spectrum for MI visibility by defining the difference in the visibilities $\Delta$, which corresponds to the band defined through two Red curves in Fig. 2.
As  a function of external modulation frequency $f$ (Hz), Fig. 4 reveals the resonance spectrum for MI.
For different input signal intensities, we also normalize the visibility difference to the same value at $f = 0$.
One can see that the spectrum shown in Fig. 4 is similar to a driven damped simple harmonic oscillator. 
Nevertheless, here we disclose  a resonance spectrum in the nonlinear system through  MI visibility.
Longer and longer relaxation time constants are needed, i.e., $\Omega \tau = 0.8$, $2.2$, $4.9$,  in order to fit into the input signal intensities for nearly instantaneous  intermediate, and non-instantaneous nonlinear responses, respectively.

Before the conclusion, we want to remark that transition behaviors in the MI visibility shown in Figs. 2 (d-i) depend strongly on the initial phase of our modulation function. Moreover, a second spatial wavenumber appears when the systems are operated in the intermediate and non-instantaneous region, see the optical images in Figs. 2(f) and 2(i).
However, to give a quantitative analysis one needs to go beyond the  nonlinear non-instantaneous Schr\"{o}dinger equation given in Eq. (1).

In conclusion,  by operating the photorefractive crystals in the  non-instantaneous region, 
we report {nonlinear resonance in modulation instability} by experimental measurements and theoretical analyses based on a nonlinear non-instantaneous Schr\"{o}dinger equation.
With a periodic modulation in the external bias voltage, which acts equivalently as a modulation in the nonlinear strength, a resonance spectrum is disclosed with an enhancement in the visibility of MI at resonant frequency.
Nonlinear manifold of  a damped oscillator is demonstrated  through spontaneous optical pattern formations, which introduces a universal control for chaotic, solitary, and turbulence waves not only in optics, but also in  plasma physics and thermodynamics.


\end{document}